\documentclass[12pt]{article}

\usepackage{amsfonts}
\usepackage{amssymb}
\usepackage{amsmath}

\usepackage{graphicx}
\usepackage{latexsym}
\usepackage{color}
\usepackage[colorlinks=true]{hyperref}
\usepackage{authblk}

\usepackage{graphicx,tikz}
\usepackage{amsmath,amssymb,mathrsfs}

\usepackage{amsthm}

\newcommand{\bm}[1]{\mbox{\boldmath $#1$}}

\def\be{\begin{equation}}
\def\ee{\end{equation}}
\def\bea{\begin{eqnarray}}
\def\eea{\end{eqnarray}}
\def\bean{\begin{eqnarray*}}
\def\eean{\end{eqnarray*}}

\def\square{\hfill\hbox{\vrule\vbox{\hrule\phantom{N}\hrule}\vrule}\,}

\newtheorem{theorem}{Theorem}[section]
\newtheorem{result}{Result}[section]
\newtheorem{coro}{Corollary}[section]
\newtheorem{defi}{Definition}
{\theoremstyle{definition}
\newtheorem{remark}{Remark}[section]}

\def\proof{\noindent{\em Proof.\/}\hspace{3mm}}

\def\scri{\mathscr{J}}

\newlength{\cellwidth}
\begin{document}
\hfill YITP-23-01
\title{Beyond black holes: Universal properties of `ultra-massive' spacetimes}

\author{Jos\'e M. M. Senovilla$^{1,2,3}$}
\affil{$^1$Departamento de F\'isica, Universidad del Pa\'is Vasco UPV/EHU, Apartado 644, 48080 Bilbao, Spain\\
$^2$Yukawa Institute for Theoretical Physics,
Kyoto University, 606-8502, Kyoto, Japan.\\
$^3$ EHU Quantum Center, Universidad del Pa\'{\i}s Vasco UPV/EHU.}

{\let\newpage\relax\maketitle}

\vspace{-0.2em}

\begin{abstract}
It has been long known that in spacetimes with a positive cosmological constant $\Lambda >0$ the area of spatially stable marginally trapped surfaces (MTSs) has a finite upper bound given by $4\pi/\Lambda$. In this paper I show that any such spacetime containing spatially stable MTSs with area approaching indefinitely that bound acquire universal properties generically. 
Specifically, they all possess {\em generalized} `holographic screens'  (i.e. marginally trapped tubes) foliated by MTSs of spherical topology, composed of a dynamical horizon portion and a timelike membrane portion that meet at a distinguished round sphere $\bar S$ with constant Gaussian curvature $\mathcal{K} =\Lambda$ ---and thus of maximal area $4\pi/\Lambda$. All future (past) generalized holographic screens containing $\bar S$ change signature at $\bar S$, and all of them continue towards the past (future) with non-decreasing (non-increasing) area of their MTSs. A future (past) singularity obtains. For the future case, these `ultra-massive spacetimes'  \cite{Snew} may be more powerful than black holes, as they can overcome the repulsive $\Lambda$-force and render the spacetime as a collapsing universe without event horizon enclosing those generalized holographic screens.  It is remarkable that these behaviours do not arise if $\Lambda$ is non-positive. The results have radical implications on black hole mergers and on very compact objects accreting mass from their surroundings---if $\Lambda >0$.
\end{abstract}

\section{Introduction}
The measurements showing that the observable Universe is in accelerated expansion \cite{Riess1998,Perlmutter1999} demand a full analysis of the implications that a {\em positive} cosmological constant $\Lambda >0$ has. One particular known result is that $\Lambda \neq 0$ imposes a finite bound
\be\label{lim}
\Lambda A < 4\pi (1-g)
\ee
on the area $A$ of marginally future-trapped\footnote{\label{foot1}All the results are applicable, {\em mutatis mutandis}, to marginally past-trapped surfaces, but for the sake of conciseness I will from the start just consider future-trapped surfaces.} closed surfaces of genus $g$ \cite{HSN,W,Simon} if these are stable in spatial directions in the sense of \cite{AMS,AMS1}---equivalently, `outer' in the sense of \cite{Hay,HSN}---and the dominant energy condition holds. The stability assumption is usually interpreted \cite{AG,HSN,SILS,Hay,PBH,BBGV} as implying that the marginally trapped surfaces (MTSs) enclose a black hole (BH) region, leading to a maximum size of BHs if $\Lambda >0$---in which case the above bound contains information only if $g=0$, topological spheres. 

To understand these physical limitations, I recently \cite{Snew} considered some simple, spherically symmetric, models where the area bound is surpassed. In spherical symmetry and for fixed $\Lambda$, the area of marginally trapped round spheres depends on the total mass enclosed by the surface, and increases as the mass also does. Therefore, one may wonder what happens if a BH in formation with area very close to the bound \eqref{lim} accretes, or receives, extra mass from its surroundings.  As shown with explicit models in \cite{Snew}, the dynamical horizon foliated by marginally trapped round spheres merges with a timelike membrane at a distinguished round sphere $\bar S$ with area $A_{\bar S}=4\pi/\Lambda$ forming a marginally trapped tube \cite{AG,Booth,BBGV} that changes signature precisely at $\bar S$. The entire marginally trapped tube (MTT) foliated by MTSs becomes locally a {\em holographic screen} as introduced in \cite{BE,BE1}, but globally is more general than that as it may contain null portions that are non-expanding horizons\footnote{These null portions are actually extremely important as they describe the periods where a BH or ultra-massive spacetime in formation is in equilibrium. These periods may have very large durations.}. Still, these MTTs satisfy an area law, for the area is always non-decreasing and strictly increasing outside the non-expanding horizon portions, and thus I will call them {\it generalized holographic screens} (GHS). The compact objects creating the intense gravitational field were destined to become BHs, but eventually something {\em more powerful} emerges and future infinity fully disappears: instead a universal future singularity arises. The spacetime becomes a contracting universe everywhere outside the spherical GHS.  I called these spacetimes `ultra massive' because the area of MTSs goes {\it beyond} the maximal bound \eqref{lim}.

In this paper I prove that the above properties are actually universal, and {\em generically} all spacetimes in which the area bound \eqref{lim} is approached without limit by spatially stable MTSs develop GHSs with $\mathbb{R}\times \mathbb{S}^2$ topology, satisfying area laws and containing a distinguished MTS $\bar S$ that is stable in a null direction, with constant Gaussian curvature ${\cal K}=\Lambda$---and thus of maximal area $A_{\bar S}=4\pi/\Lambda$. All possible marginally trapped tubes passing through $\bar S$ change signature somewhere on $\bar S$ and are GHSs. A future singularity arises, and no BH is left over but rather an ultra-massive spacetime.\footnote{The only exception to the generic situation is given by spacetimes that are (locally) foliated by non-expanding horizons \cite{AK} including the family of near-horizon geometries that contain multiple Killing horizons with compact sections \cite{MPS,MPS1,MPS2,PLJ,LS,LSW}.}
To that end, I use the properties of the stability operator for MTSs introduced in \cite{AMS,AMS1} following pioneering work by Newman \cite{Ne}. I collect the necessary (known and new) results in section \ref{sec:L}. The main results are proven in section \ref{sec:main} for the generic situation. 

The paper ends with a brief discussion in section \ref{sec:discussion}. A particularly important consequence of these results is that, at least in some cases, the merging of very big BHs may not lead to a larger BH but rather to an ultra-massive spacetime. The amount of mass required for this was estimated in \cite{Snew}, for spherically symmetric collapses, by using the present accepted value of $\Lambda \simeq 1.1 \times 10^{-52}$m$^{-2}$ based on the accelerated-expansion observations. If this estimation is approximately valid in general, at least $10^{8}$ big galaxies as massive as, say, M81 would be needed to collide and form the ultra-massive spacetime. Or equivalently, at least $10^{10}$ milky-ways. The total number of estimated galaxies in the observable Universe is about $10^{11}$, but the real number is probably larger.

But before anything, let me fix the set up and the main concepts to be used.

\subsection{Terminology and definitions}
$(M,g)$ denotes a 4-dimensional oriented Lorentzian manifold with causally oriented metric $g_{\mu\nu}$ of signature $(-,+,+,+)$ that satisfies Einstein field equations with cosmological constant $\Lambda$
\be\label{efe}
G_{\mu\nu}+\Lambda g_{\mu\nu} = \frac{8\pi G}{c^4} T_{\mu\nu}:={\cal T}_{\mu\nu} 
\ee
where $G_{\mu\nu}$ is the Einstein tensor and $T_{\mu\nu}$ an energy-momentum tensor satisfying the dominant energy condition (DEC). For the purposes of this work, a closed spacelike surface $S$ is any embedded and orientable 2-dimensional connected and compact manifold with positive-definite first fundamental form $h_{AB}$\footnote{Greek indices are spacetime indices and Latin upper-case indices belong to the 2-dimensional $S$}. I denote by $\mathfrak{X}(S)$ and $\mathfrak{X}^\perp(S)$ the set of vector fields tangent  and orthogonal to $S$, respectively. One can choose a couple of {\em null} vector fields in $\mathfrak{X}^\perp(S)$, denoted by $\vec \ell$ and $\vec k$, that are linearly independent all over $S$ \cite{Mars}. For convenience, I will choose them to be future pointing and normalized such that
$$
\ell_\mu k^\mu =-1.
$$

Let $\vec v,\vec w\in \mathfrak{X}(S)$, then one can split the covariant derivative $\nabla_{\vec v} \vec w$ into tangent and normal parts
$$
\nabla_{\vec v} \vec w =D_{\vec v} \vec w -\vec{K}(\vec v,\vec w)
$$
where $D$ is the intrinsic covariant derivative in $(S,h)$ and $\vec K$ is the shape tensor, also known as the second fundamental form vector \cite{O,Kri,S}. It can be decomposed into its trace and traceless parts relative to $h_{AB}$ as follows
$$
\vec{K}_{AB} = \vec\Sigma_{AB} +\frac{1}{2} \vec H h_{AB}, \hspace{1cm} h^{AB} \vec\Sigma_{AB} =0.
$$
Here $\vec\Sigma_{AB}$ is the total shear tensor \cite{CSV} and $\vec H$ is the mean curvature vector of $S$ in $M$ \cite{Kri,O,S}.
By definition, $\vec{K}(\vec v,\vec w), \vec\Sigma (\vec v, \vec w)$ and $\vec H$ are normal to $S$ and thus they can be expressed in the basis $\{\vec\ell,\vec k\}$:
$$
\vec K_{AB} =-K^\ell_{AB} \vec k-K^k_{AB} \vec\ell , \hspace{1cm} \vec\Sigma_{AB} =-\Sigma^\ell_{AB} \vec k-\Sigma^k_{AB} \vec\ell 
$$
and 
\be\label{H}
\vec H=-\theta^\ell \vec k -\theta^k \vec \ell
\ee
where $K^\ell_{AB} := \ell_\mu K^\mu_{AB}$ and $K^k_{AB}:=k_\mu K^\mu_{AB}$ are the null second fundamental forms, $\Sigma^\ell_{AB} := \ell_\mu \Sigma^\mu_{AB}$ and $\Sigma^k_{AB}:=k_\mu \Sigma^\mu_{AB}$ the null shear tensors, and $\theta^\ell :=\ell_\mu H^\mu$ and $\theta^k:=k_\mu H^\mu$ the null expansions of $S$ relative to $\vec\ell$ and $\vec k$ respectively.

The causal character of $\vec H$ leads to a classification into different types of surfaces, see e.g. \cite{AG,AMS,AMS1,MaSe,S,S0}. For our purposes, we just need to know the following (recall footnote \ref{foot1}): $S$ is said to be
\begin{itemize}
\item marginally `outer' trapped (MOTS) if $\vec H$ is null everywhere on $S$ consistently pointing along one of the null normals
\item marginally future trapped (MTS) if it is a MOTS with $\vec H$ future pointing all over $S$
\item weakly future trapped (WTS) if $\vec H$ is future pointing everywhere on $S$ 
\item future trapped (TS) if $\vec H$ is timelike and future pointing everywhere on $S$ 
\item (fully) untrapped if $\vec H$ is spacelike somewhere (everywhere) on $S$
\item Extremal if $\vec H$ vanishes identically on all of $S$
\end{itemize}
Equivalently, MOTSs have one of the null expansions vanishing, MTSs have in addition the other null expansion negative, WTSs and TSs have both null expansions non-positive and negative, respectively. In (fully) untrapped surfaces the expansions have opposite signs somewhere (everywhere) on $S$.

For MOTSs and MTSs the mean curvature vector points along one of the null normals ---the one with vanishing expansion---, and this is usually called the `outer' direction in the literature. But beware, `outer' has no outward meaning in principle. To fix the notation, for MOTSs and MTSs I am going to assume that $\vec k$ is the direction of the mean curvature vector, so that in this paper we are going to have
\be\label{H}
\mbox{MOTS}: \vec H = -\theta^\ell \vec k, \hspace{1cm} \mbox{MTS}: \vec H = -\theta^\ell \vec k \, \, \, \& \, \, \, \theta^\ell <0.
\ee
Notice that any MTS is, in particular, a MOTS.

A marginally (outer) trapped tube M(O)TT is a hypersurface in $M$ foliated by M(O)TSs. A spacelike MTT is usually called a dynamical horizon \cite{AK,AG} or future outer trapping horizon \cite{Hay}, while if it is timelike is called a timelike  membrane \cite{AG,BBGV} or future inner trapping horizon \cite{Hay}.

\subsection{Spacetimes containing a stable marginally trapped surface}\label{subsec:possibilities}
A pioneering notion of stability of M(O)TSs was put forward by Newman \cite{Ne} and was used in \cite{Hay} to prove many relevant and useful results. An important advancement was the refinement performed in \cite{AMS,AMS1}, where for the first time an infinitesimal stability notion was found and the `stability operators' --to be defined later-- identified. See also an interesting discussion in \cite{PBH}. I am going to make extensive use of those results and the underlying ideas. 

The basic notion is the {\em stability} of a MOTS along a given normal direction $\vec n\in \mathfrak{X}^\perp(S)$ with $n_\mu k^\mu \geq 0$ ---I will call these directions `external'. One can visualize such directions by imagining the local null hypersurface  generated by $\vec k$, and then $\vec n$ points towards its exterior, or along it if $\vec n =\vec k$, see also figure \ref{fig:scheme} below. 
\begin{defi}
A MOTS $S$ is said to be stable along an external $\vec n$ if there exists a non-negative function $f\not\equiv 0$ such that the variation of the vanishing expansion along $f\vec n$, denoted by $\delta_{f\vec n} \theta^k $, is non-negative. And it is called strictly stable if in addition $\delta_{f\vec n} \theta^k >0$ somewhere.
\end{defi}

To understand this nomenclature and the underlying conceptual scheme, note that if a MTS is embedded in a hypersurface $\Sigma$ and is strictly stable along the direction tangent to $\Sigma$ and orthogonal to $S$, then no nearby WTS extends to the `exterior' of $S$, and no nearby fully untrapped surface enters the interior of $S$, in $\Sigma$ \cite{AMS,AMS1} --recall that here exterior is defined by the direction of the null mean curvature vector. This is a kind of barrier property separating, within $\Sigma$, trapped from untrapped surfaces. Furthermore, if a MTS $S$ has this type of barrier property in $\Sigma$, then it is stable within $\Sigma$ \cite{AMS,AMS1}. It is in this sense that MTSs that are stable in spacelike directions are supposed to reside in local-in-time black-hole regions.

A deeper analysis of this stability concept will be implicit later in section \ref{sec:L}. For the present introductory purposes, let me just recall two fundamental results proven in \cite{AMS1}: 
\begin{theorem}\label{th1}
Let $S_0$ be a MOTS strictly stable along a exterior-pointing normal direction $\vec n$. Choose a hypersurface $\Sigma_0$ tangent to $\vec n$ at $S_0$ such that $S_0$ is embedded in $\Sigma_0$, and a family of hypersurfaces $\{\Sigma_s\}$, $s\in[0,T]$, foliating the spacetime locally near $S_0$. Then, for some $\tau\in (0,T]$ there is a local MOTT adapted to the foliation, in the sense that each MOTS $S_s$ of the MOTT is embedded in $\Sigma_s$ for all $s\in [0,\tau)$ and strictly stable there. In addition, the MOTT thus built is never tangent to any of the $\Sigma_s$.
\end{theorem}
Obviously, the MOTT depends on the choice of reference foliation $\{\Sigma_s\}$. Notice moreover that, as long as the $S_s$ remain smooth and embedded, the `evolution' of the MOTT continues until either the MOTSs run out to infinity, or strict stability is lost. In the latter case we also have
\begin{theorem}\label{th2}
Under the same hypothesis of theorem \ref{th1}, assume that the MOTSs $\{S_s\}$ converge  as $s\rightarrow \tau$ to a smooth closed MOTS $S_\tau$ that is stable, but not strictly stable, in $\Sigma_\tau$. If the function
\be\label{W}
W:= {\cal T}_{\mu\nu}k^\mu k^\nu + \Sigma^k_{AB} \Sigma^k{}^{AB}
\ee
is not identically zero on $S_\tau$, then the MOTT extends to all $s\in [0,\tau]$ and is tangent to $\Sigma_\tau$ at $S_\tau$.
\end{theorem}
Observe that the function $W$ is always non-negative (for this DEC is not needed, the mere null convergence condition is enough). This function $W$ plays an important role in what follows, as the possibility $W\equiv 0$ determines the exceptional cases.
When the reference foliation is composed by {\em non-timelike} hypersurfaces then the MOTT of the above theorems is spacelike everywhere near $S_0$ if $W|_{S_0} \not\equiv 0$, and is null on $S_0$ if $W$ vanishes identically there.

Therefore, every spacetime satisfying DEC and containing a MTS $S_0$ that is strictly stable in some {\em spacelike} external direction contains MTTs that are dynamical horizons around $S_0$ if $W|_{S_0} \not\equiv 0$ and must fall into one of the following categories:
\begin{enumerate}
\item All MTTs containing $S_0$ eventually become simple MOTTs--or reach an extremal surface--loosing the negativity of the non-vanishing expansion of the foliating MOTSs, or the M(O)TSs develop singularities, or in general something happens that stops the `evolution' of these MTTs.
\item evolution of (at least) some MTT continues indefinitely but the foliating MTSs have areas that never approach the bound \eqref{lim}. These include standard situations where, if there is a well defined future infinity $\scri^+$ \cite{P00,HE,Wald,Fra}, a global event horizon encloses a BH and the MTT, and this either merges with, or asymptotically approaches, the global event horizon ---whose area is also bound by \eqref{lim}.
\item evolution of (at least) some MTT continues such that the foliating MTSs have areas that approach the bound \eqref{lim} as much as desired, but the entire MTT can never be contained in a compact set. These are marginal cases where the bound is never attained but is `reached' somehow at infinity. A particular explicit example of this behaviour is given by the toy model of figure 4 in \cite{Snew}, but note that in this example infinity has a very peculiar structure, and is reachable only by some special observers.
\item \label{pos1} evolution of (at least) some MTT continues such that the foliating MTSs have areas that approach the bound \eqref{lim} as much as desired, and the entire MTT is contained in a compact set. Several explicit examples of this case were presented in \cite{Snew}.
\item \label{pos2} there are finally the exceptional cases where the function $W\equiv 0$ for a set of MTSs that cover an open region of the spacetime and of the MTTs.
\end{enumerate}
Possibility \ref{pos1} is the one I wish to analyze in this work. This will be done in section \ref{sec:main} where I also provide a precise formulation of the type of spacetimes to be considered---ultra-massive spacetimes---in Definition \ref{ultraM}. For that, I will need some fundamental results on MTSs and MTTs that are presented in the next section. The exceptional situation \ref{pos2} will hopefully be studied in later work.

\section{The stability operator for MOTSs at work}\label{sec:L}
In this section I collect all results on MTSs and MTTs that I will need in what follows. Many of them are known and can be found in \cite{AMS,AMS1,AK,AG,BeS,BBGV,BE,BE1,Hay,JRD,Mars,Ne,PBH,SPrague,SERE,SILS,Simon}, some others are new as far as I know. The basic tool I am going to use is the {\em stability operator} for MOTSs, as introduced in \cite{AMS,AMS1}.

\begin{figure}[h!]
\includegraphics[width=12cm]{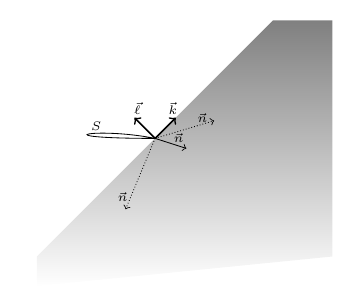}
\caption{Scheme of the directions for a M(O)TS $S$. $\vec k$ and $\vec\ell$ are the two future null directions normal to $S$ and $\vec n \in \mathfrak{X}^\perp (S)$ is any possible direction pointing towards the shadowed zone, which is the {\em exterior} region for $S$. The choice of normalization is such that $k_\mu n^\mu =1$, so that the norm $n_\mu n^\mu$ of $\vec n$ chooses the particular direction. Of course, the direction of $\vec n$ depends on the point on $S$. Observe that the causal character of $\vec n$ is unrestricted and that $\vec k$ is the only normal direction not included in the set of vector fields $\vec n$.\label{fig:scheme}}
\end{figure}

Let $S$ be a MOTS as explained before. The external pointing vector fields previously introduced are going to be characterized by their norms as follows, see the pictorial explanation in Figure \ref{fig:scheme}
\be
\vec n =-\vec\ell +\frac{n_{\mu}n^{\mu}}{2}\vec k  \label{n}
\ee
so that the normalization with respect to the mean curvature vector is
\be
H_\mu n^\mu = \theta^\ell \hspace{1cm} (k_\mu n^\mu=1) .\label{norm}
\ee
The causal character of $\vec n$ is unrestricted. Let $\vec u\in \mathfrak{X}^\perp(S)$ be the `Hodge dual' of $\vec n$ in the normal bundle, that is, the unique vector field normal to $S$, orthogonal to $\vec n$ and with opposite norm:
\be
\vec u = \vec\ell +\frac{n_{\mu}n^{\mu}}{2}\vec k, \hspace{1cm} u_\mu n^\mu =0, \hspace{1cm} u_\mu u^\mu =-n_\mu n^\mu  .\label{u}
\ee

The formula for the variation of the vanishing expansion $\theta^k=0$ reads
\be
\delta_{f\vec n} \theta^k:=L_n f
\label{deltatheta}
\ee
where $L_n$ is the {\em stability operator} for $S$ in the direction $\vec n$ explicitly given by  \cite{AMS,AMS1}
\be
L_n f =-\Delta f+2s^BD_{B}f+f\left({\cal K} -s^B s_{B}+D_{B}s^B-\frac{n^\rho n_{\rho}}{2}\Sigma^k_{AB} \Sigma^k{}^{AB}-G_{\mu\nu}k^\mu u^{\nu} \right)
\label{Ln0}
\ee
where ${\cal K}$ is the Gaussian curvature on $S$, $\Delta$ its Laplacian, and $s_{B}$ the one-form on $S$ defined by
$$
\bm{s} (\vec v) =k_{\mu}\nabla_{\vec v}\ell^\mu, \hspace{1cm} \forall \vec v\in \mathfrak{X}(S) .
$$
Using the definiton of $W$ in \eqref{W}, an alternative form of $L_n$ is
\be
L_n f =-\Delta f+2s^B D_{B}f+f\left({\cal K} -s^B s_{B}+D_{B}s^B-G_{\mu\nu}k^\mu \ell^{\nu}-\frac{n^\rho n_{\rho}}{2}\,  W\right) \label{Ln}
\ee
For positive $f>0$, yet another alternative way of writing these formulas reads \cite{JRD,Simon}
\bea
\frac{L_n f}{f} &=& -\Delta \ln f +D_B s^B-\left(D_B \ln f -s_B\right)\left(D^B \ln f -s^B\right)+{\cal K}-\frac{n^\rho n_{\rho}}{2}\Sigma^k_{AB} \Sigma^k{}^{AB}-G_{\mu\nu}k^\mu u^{\nu}\nonumber\\
&=&-\Delta \ln f +D_B s^B-\left(D_B \ln f -s_B\right)\left(D^B \ln f -s^B\right)+{\cal K}-\frac{n^\rho n_{\rho}}{2}W-G_{\mu\nu}k^\mu \ell^{\nu} \label{Ln2}.
\eea
Formula (\ref{deltatheta}) is valid for all possible normal directions {\em except} for $\vec k$ (due to the normalization used). The variation in that distinguished null direction reads simply
\be
\delta_{f\vec k} \theta^k= -f W, \hspace{5mm} \mbox{in particular}\hspace{5mm}  \delta_{\vec H} \theta^k= \theta^\ell W . \label{deltak}
\ee
Thus, the variation of the vanishing expansion along the corresponding {\em future} null direction is non-positive if the null convergence condition holds.

For each $\vec n$, $L_n$ is an elliptic operator on $S$, not self-adjoint in general (with respect to the $L^2$-product on $S$). It possesses a {\em real} principal eigenvalue, denoted by $\lambda_n$, and the corresponding {\em real} eigenfunction, denoted by $\phi_n$, can be chosen to be positive on all of $S$. The formal adjoint operator of $L_n$ with respect to the $L^2$-product is given by 
$$
L^\dag_n f =-\Delta f-2s^B D_{B}f+f\left({\cal K} -s^B s_{B}-D_{B}s^B-\left.G_{\mu\nu}k^\mu \ell^{\nu}\right|_S-\frac{n^\rho n_{\rho}}{2}\,  W\right) 
$$
and has the same principal eigenvalue as $L_n$. The corresponding principal eigenfunctions are denoted by $\phi^\dag_n$, and they are real and positive on the entire $S$.


The first fundamental result that I am going to use is \cite{AMS,AMS1}
\begin{result}
A MOTS is (strictly) stable along the external direction $\vec n$ if and only if $\lambda_n$ is non-negative (positive). 
\end{result}
Expression \eqref{Ln} implies that the (in)stability of $S$ is independent of the direction if $W\equiv 0$, it is in this sense that the case $W\equiv 0$ is exceptional. Therefore, it is useful to introduce the following definition.
\begin{defi}\label{Wset}
For any closed surface, ${\cal W}(S) \subset S$ is defined as the set of points in $S$ with vanishing function \eqref{W}:
$$
{\cal W}(S) := \{q\in S, \hspace{2mm} W|_q =0\}.
$$
\end{defi}
Concerning the stability along the mean curvature vector, or $\vec k$, the following holds.
\begin{result}\label{k}
For a MOTS $S$, 
\begin{enumerate}
\item\label{Wnot0} if $S\setminus {\cal W}(S)\neq \emptyset$, $S$ is unstable along $\vec k$ and strictly stable along $-\vec k$.
\item\label{W=0} if $S= {\cal W}(S)$ then $S$ is stable along both $\pm\vec k$.
\end{enumerate}
\end{result} 
\proof This follows immediately from \eqref{deltak}.

In fact, it is easily seen that the relation between stability operators along different directions $\vec n$ and $\vec n'$ is given simply by
\be\label{twoL}
L_{n'} = L_n +\frac{W}{2} \left(n_\mu n^\mu -n'_\mu n'^\mu\right)
\ee
making it plain that the (in)stability of $S$ is independent of $\vec n\neq \vec k$ when $W\equiv 0$. Therefore, the exceptional case with $W=0$ can be classified in three possibilities.
\begin{result}\label{excep}
Let $S$ be a MOTS with $S= {\cal W}(S)$, that is, with $W=0$ on all of $S$. Then only one of the following three cases can occur:
\begin{enumerate}
\item\label{allunsta} $S$ is unstable in all external directions and stable along $\pm\vec k$.
\item\label{allssta} $S$ is strictly stable in all external directions and stable along $\pm\vec k$
\item\label{allsta} $S$ is merely stable along all external directions including $\pm\vec k$.
\end{enumerate}
\end{result}
\proof The statement about $\pm\vec k$ is point \ref{W=0} in Result \ref{k}. From \eqref{twoL} we know that the stability operator is the same for all possible external directions (different from $\pm\vec k$) and actually equal to $L_{-\ell}$. Thus, depending on the the sign of $\lambda _{-\ell}(=\lambda_n$ for all $\vec n$) the three possibilities arise.\square

These three cases arise when $S$ belongs to a non-expanding horizon (NEH), in particular for isolated horizons and Killing horizons \cite{AK}. Case \ref{allssta} is for black-hole type such horizons, while case \ref{allunsta} arises in situations such as the past cosmological horizon in de Sitter spacetime. The case \ref{allsta} is the truly exceptional possibility, and arises in regions of the spacetime foliated by NEHs, \cite{LS,LSW,PLJ} that include the multiple Killing horizons of near-horizon geometries \cite{MPS,MPS1,MPS2}. 

From formula \eqref{twoL} the following relation can be derived \cite{AMS1}
$$
(\lambda_n -\lambda_{n'}) \int_S \phi^\dag_n \phi_{n'} = -\int_S \phi^\dag_n \phi_{n'} \frac{W}{2} \left(n_\mu n^\mu -n'_\mu n'^\mu\right)
$$
 so that taking into account the positivity of the principal eigenfunctions one obtains the following set of results
\begin{result}\label{twon}
Assume that $S\setminus {\cal W}(S)\neq \emptyset$ on a MOTS $S$. Given any two normal (exterior pointing and normalized according to \eqref{norm}) directions $\vec n$ and $\vec n'$, the following list of results hold \cite{AMS1,JRD,Mars,SPrague,SERE}:
\begin{enumerate}
\item $\lambda_n -\lambda_{n'} +W\left(n_\mu n^\mu -n'_\mu n'^\mu\right)/2$ necessarily changes sign on $S$, or is identically zero;
\item\label{W=0} if $\vec n$ and $\vec n'$ differ exclusively within the region ${\cal W}(S)$, then $\lambda_n =\lambda_{n'}$;
\item\label{interweave} any two directions with the same principal eigenvalue $\lambda_n = \lambda_{n'}$ are such that $n_\mu n^\mu -n'_\mu n'^\mu$ either changes sign on $S$, or is non-zero only within ${\cal W}(S)$.
In the former case, $\vec n -\vec n'\propto \vec k$ necessarily changes sign, that is, causal orientation, on $S$;
\item\label{SupInf} $\lambda_{n'} + {\rm inf}_S \left[W\left(n'_\mu n'^\mu -n_\mu n^\mu\right)/2\right] \leq \lambda_n \leq \lambda_{n'} + {\rm sup}_S \left[W\left(n'_\mu n'^\mu -n_\mu n^\mu\right)/2\right]$
\item\label{l'>l} if $n'_\mu n'^\mu \leq n_\mu n^\mu $ everywhere outside ${\cal W}(S)$ and $n'_\mu n'^\mu < n_\mu n^\mu $ somewhere there, then $\lambda_{n'} > \lambda_n$;
\item\label{l'<l} if $\lambda_{n'} < \lambda_n$, then $n'_\mu n'^\mu > n_\mu n^\mu$ somewhere on $S\setminus {\cal W}(S)$.
\item if $S$ is not stable along the null direction $-\vec \ell$, then it cannot be stable along any spacelike direction.
\end{enumerate}
\end{result}
The last point implies that a necessary condition for the stability of MOTSs in spacelike external directions is its stability along the null direction $-\vec \ell$. Observe also that if a MOTS is stable along a spacelike $\vec n$, then it must be strictly stable along $-\vec\ell$. 
\footnote{This also implies that the notion of {\em outer trapping surface} as defined by Hayward ---not to be confused with outer trapped surfaces!--- in \cite{Hay} is essentially equivalent to that of stability along some spacelike directions.}

Plugging $f=\phi_n$ into \eqref{Ln2} and using \eqref{efe} one gets
\bea
\lambda_n &=& -\Delta \ln \phi_n +D_B s^B-\left(D_B \ln \phi_n -s_B\right)\left(D^B \ln \phi_n -s^B\right)+{\cal K}-\Lambda- \frac{n^\rho n_{\rho}}{2}\Sigma^k_{AB} \Sigma^k{}^{AB}-{\cal T}_{\mu\nu}k^\mu u^{\nu}\nonumber\\
&=&-\Delta \ln \phi_n +D_B s^B-\left(D_B \ln \phi_n -s_B\right)\left(D^B \ln \phi_n -s^B\right)+{\cal K}-\Lambda -\frac{n^\rho n_{\rho}}{2}W-{\cal T}_{\mu\nu}k^\mu \ell^{\nu}\label{lambdan}
\eea
so that integrating this on the compact $S$ and using the Gauss-Bonnet theorem
\bea
\left(\Lambda +\lambda_n \right)A_S &=&4\pi (1-g) -\int_S\left[\left(D_B \ln \phi_n -s_B\right)\left(D^B \ln \phi_n -s^B\right)+ \frac{n^\rho n_{\rho}}{2}\Sigma^k_{AB} \Sigma^k{}^{AB}+{\cal T}_{\mu\nu}k^\mu u^{\nu}\right]\nonumber \\
&=&4\pi (1-g) -\int_S\left[\left(D_B \ln \phi_n -s_B\right)\left(D^B \ln \phi_n -s^B\right)+ \frac{n^\rho n_{\rho}}{2}W+{\cal T}_{\mu\nu}k^\mu \ell^{\nu}\right]\label{genineq}
\eea
where $A_S$ is the area of $S$ and $g$ its genus. Therefore, if the DEC holds and the normal direction $\vec n$ is non-timelike everywhere, the integrand on the righthand side is non-positive and one gets the bound \eqref{lim} together with some other consequences \cite{GSc,HSN,JRD,Mars,Simon,W}. 
\begin{result}\label{areas}
Assume that DEC holds and $\Lambda \neq 0$
in a spacetime that satisfies the Einstein field equations \eqref{efe} and let $S$ be a MOTS with $S\setminus {\cal W}(S)\neq \emptyset$, ${\cal W}(S)$ as introduced in definition \ref{Wset}. Then
\begin{enumerate}
\item \label{ineq}For any external direction $\vec n$ that is non-timelike on $S\setminus {\cal W}(S)$, the following inequality holds
$$
\left(\Lambda +\lambda_n \right)A_S\leq 4\pi (1-g) 
$$
with equality possible only if
\be\label{zeros}
s_B =D_B \ln \phi_n, \hspace{1cm} (n_\mu n^\mu) \Sigma^k_{AB} \Sigma^k{}^{AB}= 0, \hspace{1cm} T_{\mu\nu} k^\mu u^\nu=0
\ee
\item\label{ineq1} In particular, if $S$ is stable ($\lambda_n\geq 0$) along an external $\vec n$ that is non-timelike on $S\setminus {\cal W}(S)$ then
$$
\Lambda A_S\leq 4\pi (1-g) 
$$
equality requiring $\lambda_n=0$ and \eqref{zeros}
\item\label{topol} $S$ cannot be stable in external directions that are non-timelike on $S\setminus {\cal W}(S)$ if $\Lambda >0$ and $g>0$.
\item\label{Alarger} If $\Lambda >0$ and the area $A_S > 4\pi /\Lambda$, then $\lambda_n <0$ necessarily along all external $\vec n$ that are non-timelike on $S\setminus {\cal W}(S)$. In particular $\lambda_{-\ell} <0$.
\item\label{Aequal} If $\Lambda >0$ and $A_S= 4\pi /\Lambda $, then $\lambda_n \leq 0$ in all external directions that are non-timelike on $S\setminus {\cal W}(S)$.
\item\label{Aequal1} If $\Lambda >0$, $A_S= 4\pi/\Lambda $ and $\lambda_n =0$ along an external $\vec n$ that is non-timelike on $S\setminus {\cal W}(S)$, then \eqref{zeros} hold and $S$ is a metric round sphere of constant curvature ${\cal K}=\Lambda$. In this case, the energy-momentum tensor must take the following form on $S$:
\be\label{emt}
T_{\mu\nu}|_S = a^2 k_\mu k_\nu + b^2 \ell_\mu \ell_\nu +C_{\mu\nu}
\ee
where $C_{\mu\nu}$ is a symmetric tensor field tangent to $S$ ($k^\mu C_{\mu\nu}=0=\ell^\mu C_{\mu\nu}$) such that
\be\label{dec}
a^2b^2 \geq C_{\mu\rho} C_{\nu}{}^\rho v^\mu v^\nu
\ee
for {\em all} unit vectors $\vec v \in \mathfrak{X}(S)$. At the points where $\vec n$ is spacelike ($n_\mu n^\mu >0$ there) one further has $b=0$ and $C_{\mu\nu}=0$. Moreover, $b=0$ and $C_{\mu\nu}=0$ on ${\cal W}(S)$.
\end{enumerate}
\end{result}
\begin{remark}\label{rem:Tkl}
Note that the last in \eqref{zeros}  always includes $T_{\mu\nu}k^\mu \ell^\nu =0$ due to DEC.
\end{remark}

\proof Statements \ref{ineq} through \ref{Aequal} are direct consequences of \eqref{genineq}. 
To prove \ref{Aequal1}, first note that \eqref{zeros} hold due to statement \ref{ineq1}. Introducing \eqref{zeros} and $\lambda_n=0$ into \eqref{lambdan} one has
$$
0= -\Delta \ln\phi_n +\Delta \ln \phi_n +{\cal K} -\Lambda
$$
so that the Gaussian curvature of $S$ is constant and positive, ${\cal K}=\Lambda$ (and thus $g=0$). To deduce the form \eqref{emt} for $T_{\mu\nu}$ I repeatedly use the properties of tensors satisfying the dominant property, see e.g. \cite{BS}. As $u^\mu$ is future, $T_{\mu\nu}u^\nu$ must be past directed, and the last in \eqref{zeros} readily implies $T_{\mu\nu}u^\nu=-a^2 k_\mu$ for some $a$. Similarly, $T_{\mu\nu}k^\mu$ is past directed, so that the last in \eqref{zeros} again provides $T_{\mu\nu}k^\nu =-b^2 \ell_\mu$, and moreover $b$ must vanish at any point where $\vec u$ is timelike, that is, at any point where $\vec n $ is spacelike: $b^2 n_\mu n^\mu =0$. The structure \eqref{emt} follows. Take then any other future null direction $\vec M$ on $S$, that is 
$$
\vec M = \vec v + x\vec k + y\vec\ell, \hspace{1cm} v_\mu v^\mu =1, \hspace{1cm} 2xy =1.
$$
Then, DEC states that $T_{\mu\nu} M^\nu = -a^2 y k_\mu -b^2 x \ell_\mu +C_{\mu\nu} v^\nu$ is past directed, and its non-positive norm leads to \eqref{dec}. On ${\cal W}(S)$ $T_{\mu\nu}k^\mu k^\nu =0$, and thus $b=0$ follows too. \square

Let us now consider MOTTs ${\cal H}$ containing a given MOTS $S$ with $S\setminus {\cal W}(S) \neq \emptyset$. As the MOTS $S$ is embedded in ${\cal H}$, taking appropriate deformations of $S$ along the direction tangent to the MOTT and normal to $S$, say $\vec m\in \mathfrak{X}^\perp(S) \cap \mathfrak{X}(\cal H)$, will lead to another MOTS of the MOTT. Hence, along this direction $\lambda_m=0$, and the deformation leading to another MOTS is given by the vector field $\phi_m \vec m$. In other words, the possible directions of MOTTs containing a given MOTS $S$ with $S\setminus {\cal W}(S) \neq \emptyset$ are included in the set of directions with vanishing eigenvalue. Then, point \ref{interweave} in Result \ref{twon} immediately provides the next result \cite{AG,BeS,SPrague,SERE}.
\begin{result}\label{inter}
Any two MOTTs containing the same MOTS $S$ (such that $W|_S \not \equiv 0$) necessarily interweave each other near $S$ in such a way that their tangent vector fields orthogonal to the MOTS and subject to \eqref{norm}, say $\vec m$ and $\vec m'$, are such that $\vec m -\vec m' \propto \vec k$ changes causal orientation on $S$ or is only non-vanishing within ${\cal W}(S)$.
\end{result}

\subsection{Existence of (in)stability directions}
The number of directions with vanishing eigenvalue is huge \cite{SPrague,SERE}, and they contain many possible MOTTs through the same MOTS $S$ as long as $S$ is stable is some external direction $\vec n$. The existence of these many MOTTs through $S$  follows from Theorem \ref{th1} by changing the reference foliation at will. The question of whether or not a given MOTS can be unstable in {\em all} possible external directions can be easily answered in the negative if $W\not\equiv 0$ \cite{Mars,AMS1}
\begin{result}\label{always}
Assume the null convergence condition holds and let $S$ be a MOTS with $S\setminus {\cal W}(S) \neq \emptyset$. There always exist external stability directions, as well as external directions into which $S$ is strictly stable and external directions into which $S$ is unstable.
\end{result}
\proof From point \ref{SupInf} in Result \ref{twon} we know that
$$
\lambda_n \leq {\rm sup}_S \left[W(n'_\mu n'^\mu -n_\mu n^\mu)/2 \right] +\lambda_n'
$$
so that choosing $\vec n$ sufficiently close to $\vec k$, that is, choosing $n_\mu n^\mu$ large enough and as $W>0$ somewhere on $S$, the righthand side can be made negative, and thus $\lambda_n <0$ for such a choice. Similarly, using now the other inequality side in the same point \ref{SupInf} of Result \ref{twon} and choosing $n_\mu n^\mu$ negative enough one can achieve $\lambda_n >0$ for some $\vec n$. To prove that there always exist directions with $\lambda_n =0$, start with a direction $\vec n'$ of strict stability, and use Theorem \ref{th1} to construct an adapted MOTT containing $S$. As explained above the direction $\vec m$ tangent to the MOTT and orthogonal to $S$ has a vanishing variation $\delta_{\phi_m \vec m} \theta^k =0$ of the vanishing expansion  and thus $\lambda_m=0$.\square

Actually, many (in)stability directions can be explicitly identified in the case that $W\neq 0$ all over $S$.
\begin{result}\label{L+-}
Assume the null convergence condition holds and let $S$ be a MOTS with ${\cal W}(S) =\emptyset$. Then
\begin{enumerate}
\item\label{alwaysstab} $S$ is strictly stable along all normal directions with
$$
Wn_\mu n^\mu\leq 2\left( {\cal K}-s_B s^B + |D_B s^B| -G_{\mu\nu}k^\mu \ell^\nu\right) \hspace{5mm} \mbox{(and $<$ somewhere)}
$$
\item\label{alwaysinestab} And $S$ is unstable along all normal directions with
$$
Wn_\mu n^\mu\geq 2\left( {\cal K}-s_B s^B - |D_B s^B| -G_{\mu\nu}k^\mu \ell^\nu\right) \hspace{5mm} \mbox{(and $>$ somewhere)}
$$
\item\label{K>Lambda} If $S$ is stable along a non-timelike direction and DEC holds, then $S$ must have at least a point $q\in S$ where ${\cal K}|_q \geq \Lambda$. And if $\vec n$ is spacelike somewhere then ${\cal K}|_q >\Lambda$ is actually necessary.
\end{enumerate}
\end{result}
\proof 
In \cite{SERE,SPrague} it was proven that, if $W\neq 0$, the two external directions $\vec n_\pm$ defined by
$$
W n_\pm{}_\mu n_\pm^\mu = 2\left( {\cal K}-s_B s^B \pm D_B s^B -G_{\mu\nu}k^\mu \ell^\nu\right)
$$
have vanishing principal eigenvalues $\lambda_{n_\pm} =0$, ergo these are explicit examples of stability directions. Then, Result \ref{twon} implies that every $\vec n$ with $n_\mu n^\mu \geq  n_\pm{}_\mu n_\pm^\mu $ (and $>$ somewhere) will be unstable directions, while those with with $n_\mu n^\mu \leq n_\pm{}_\mu n_\pm^\mu $ (and $<$ somewhere) will be directions of strict stability, so that points \ref{alwaysstab} and \ref{alwaysinestab} follow. To prove point \ref{K>Lambda} assume on the contrary that ${\cal K} < \Lambda $. Then, using the field equations \eqref{efe} and point \ref{alwaysinestab} instability of $S$ along all directions $\vec n$ with
$$
Wn_\mu n^\mu\geq 2\left( {\cal K}-s_B s^B - |D_B s^B| -{\cal T}_{\mu\nu}k^\mu \ell^\nu -\Lambda \right)< 0
$$
would follow, in contradiction.
\square

The stability operators corresponding to $\vec n_\pm$ read
$$
L_{n_+}f= -\Delta f+2s^B D_B f, \hspace{1cm} L_{n_-}f =-\Delta f +2s^BD_B f +2fD_B s^B 
$$
and they have some specific interesting properties \cite{SPrague,SERE}. Notice in particular that the principal eigenfunction $\phi_{n_+}$ is a constant.

Identifying specific stability directions when ${\cal W}(S)$ is non-empty is a more difficult task.

\subsection{Area law: generalized holographic screens}
Concerning the variation of area for MOTSs foliating a MOTT, let $\epsilon_{AB}$ be the canonical area 2-form on $(S,h)$. From classical results in semi-Riemannian geometry, the variation of $\epsilon_{AB}$ along any possible direction defined by a vector field $\vec \xi$ is ruled by the mean curvature vector as follows (see e.g. \cite{Kri,O,MaSe})
$$
\delta_\xi \epsilon_{AB} = \epsilon_{AB} \left(D_B \xi^B_\parallel +H_\mu \xi^\mu \right)
$$
where $\vec\xi_\parallel$ is the part of $\vec\xi$ tangent to $S$. In particular, the total variation of area is thus
$$
\delta_\xi A_S = \int_S H_\mu \xi^\mu .
$$
Note that only the normal part of $\vec\xi$ enters in this relation. Thus, restricting to $\vec\xi\in \mathfrak{X}^\perp (S)$ and recalling \eqref{H}, the variations become
\be
\delta_\xi \epsilon_{AB} =- \epsilon_{AB} \theta^\ell (k_\mu \xi^\mu), \hspace{1cm} \delta_\xi A_S =-\int_S \theta^\ell  (k_\mu \xi^\mu).
\ee
Recall that the external vector fields $\vec n$ that we are using in the stability operator have $n_\mu k^\mu  =1$. This leads to a result, the area law, for MTTs \cite{Hay,AK,BE,BE1}.
\begin{result}\label{area}
The area element $\epsilon_{AB}$ and the total area of the MTSs foliating any MTT are everywhere non-decreasing along external directions, and can only remain stationary at the regions where the MTT is tangent to the mean curvature vector of the MTSs, that is, tangent to $\vec k$. 
\end{result}
This is the basic property of ``holographic screens'' as defined in \cite{BE,BE1}, see also \cite{SW}. In those references the authors required that $W\neq 0$ everywhere and that the stationary set was with empty interior. These assumptions are not necessary, and one can accept open portions where the MTT is a non-expanding (or isolated or Killing  \cite{AK}) horizon tangent to $\vec k$. Explicit examples are those of Figures 2, 3 and 6 in \cite{Snew}. Thus, I will use the term ``generalized holographic screens'' for these MTTs. I would like to remark that if the MTT is tangent to the mean curvature vector --and therefore to $\vec k$-- of the foliating MTSs in an open portion, then from \eqref{deltak} it follows that $W\equiv 0$ on that portion leading to the exceptional case. Observe that the area remains constant in these portions.

Let me finally recall another important result, that provides a deeper meaning to the previous one \cite{AG,BE,BE1}.
\begin{result}
The foliation by MTSs of any MTT is unique, except in open regions where the MTT is null and tangent to the mean curvature vector of the MTSs.
\end{result} 

\section{Main results: ultra massive spacetimes}\label{sec:main}
We are now ready to analyze possibility \ref{pos1} as presented in section \ref{subsec:possibilities} of the Introduction. 
%
To that end, it is necessary to provide a more rigourous definition of what Possibility \ref{pos1} requires. 

\begin{defi}[Ultra-massive spacetimes]\label{ultraM}
A spacetime will be called ultra-massive if it contains a dynamical horizon foliated by MTSs $\{S_s\}$ ($s\in [0,\tau)$) with the following properties
\begin{itemize}
\item All of them are such that the set ${\cal W}(S_s)$ is not the entire $S_s$
\item For any positive $\epsilon >0$, some of the foliating MTSs have areas larger than $4\pi/\Lambda -\epsilon$.
\end{itemize}
\end{defi}
The second condition here captures the idea that the areas of the MTSs in the dynamical horizon approach the bound \eqref{lim} indefinitely.
\begin{remark}
Observe that it would actually be sufficient to assume the existence of the MTS $S_0$ and its stability along a spacelike direction because then theorem \ref{th1} applies and, choosing the reference foliation $\{\Sigma_s\}$ composed by non-timelike hypersurfaces, the resulting MOTT will be spacelike and foliated by MTSs---as the expansion $\theta^\ell <0$ on $S_0$. Thus, the MOTT will be a dynamical horizon locally. The true assumptions in definition \ref{ultraM} are that the $\{S_s\}$ remain strictly stable along spacelike directions while at the same time their areas increase approaching the bound \eqref{lim}.
\end{remark}

\begin{theorem}
Let the spacetime be ultra-massive in the sense of definition \ref{ultraM} and assume that $T_{\mu\nu}$ satisfies the dominant energy condition and $\Lambda >0$.
Then
\begin{itemize}
\item $S_0$ belongs to a generalized holographic screen ${\cal H}$ that satisfies the area law
\item The topology of ${\cal H}$ is $\mathbb{R}\times \mathbb{S}^2$
\item There is a distinguished MTS $\bar S\in {\cal H}$ with constant Gaussian curvature ${\cal K} =\Lambda$ ---and area $4\pi /\Lambda$. 
\item $T_{\mu\nu}|_{\bar S}$ takes the form \eqref{emt}.
\item All GHSs passing through $S_0$ change signature somewhere at $\bar S$. At least one GHS is null everywhere on $\bar S$.
\item If $\bar S$ is not extremal and $\bar S\setminus {\cal W}(\bar S)\neq \emptyset$, all such GHSs become (partly) timelike towards the past of $\bar S$, and continue all along towards the past as far as the foliating surfaces keep being marginally trapped (and the genericity condition $W\not\equiv 0$ holds on them)

\end{itemize}
\end{theorem}
\proof
Given the strict stability of $S_0$ along a spacelike external direction its topology must be $\mathbb{S}^2$ (point \ref{topol} in Result \ref{areas}) and a reference foliation composed by non-timelike hypersurfaces $\{\Sigma_s\}$ can be chosen adapted to the dynamical horizon of definition \ref{ultraM} according to Theorem \ref{th1}. Thus, the topology of this DH is $\mathbb{R}\times \mathbb{S}^2$.

From theorem \ref{th2}, the DH extends to $s=\tau$ as a MTT whenever $S_\tau :=\bar S$ is not extremal, and also this MTT is tangent to $\Sigma_\tau$ everywhere on $\bar S$ if $\bar S$ is just stable (but not strictly stable) there. 
Let $\vec n_s$ be the non-timelike directions of strict stability for each $S_s$, $s\in[0,\tau)$, tangent to the hypersurfaces $\Sigma_s$. Point \ref{ineq} in Result \ref{areas} implies that 
$$(\Lambda +\lambda_{n_s})A_{S_s}\leq 4\pi, \hspace{3mm} \forall s\in [0,\tau).
$$
where $\lambda_{n_s}$ denote the principal eigenvalue of each $S_s$ along the external direction $\vec n_s$, and thus $\lambda_{n_s} >0$ for all $s\in [0,\tau)$, so that $\Lambda A_{S_s} <4\pi $  for all $s\in [0,\tau)$.
As the areas increase monotonically with $s$ due to Result \ref{area}, it follows that necessarily in the limit 
$$
A_{\bar S} = 4\pi  /\Lambda
$$
as otherwise (if $A_{\bar S} < 4\pi  /\Lambda$) there would exist some $\epsilon >0$ such that all the $S_s$ of the DH will have an area strictly smaller than $4\pi/\Lambda -\epsilon$, in contradiction.

Furthermore, as $\vec n_s$ are non-timelike for all $s\in [0,\tau)$, in the limit the stability direction $\vec n_\tau$ tangent to $\Sigma_\tau$  is non-timelike everywhere on $\bar S$. Point \ref{Aequal} in Result \ref{areas} ensures then that $\bar S$ cannot be strictly stable within $\Sigma_\tau$, and thus it has to be just stable so that at $\bar S$
$$\lambda_{n_\tau}=0$$
and the MTT is tangent to $\Sigma_\tau$ everywhere on $\bar S$.
Point \ref{Aequal1} in Result \ref{areas} then ensures ${\cal K}=\Lambda$ and the form \eqref{emt} of $T_{\mu\nu}$, plus \eqref{zeros}. Hence, if $\vec n_\tau$ is spacelike somewhere on $\bar S$, $\Sigma^k_{AB}=0$ and $W=0$ there. In other words, $\vec n_\tau$ is null and proportional to $-\vec \ell$ everywhere on $\bar S\setminus {\cal W}(\bar S)$ and, due to the construction, non-timelike on ${\cal W}(\bar S)$. This implies also that $\Sigma_\tau$ is null everywhere on $\bar S\setminus {\cal W}(\bar S)$ and, as the MTT is tangent to $\Sigma_\tau$ on $\bar S$, the MTT changes signature there. 
Moreover, using now point \ref{W=0} of Result \ref{twon} one deduces that on $\bar S$
$$\lambda_{-\ell}=0.$$
This implies that if the reference foliation is such that $\Sigma_\tau$ is null everywhere on $\bar S$, the corresponding MTT  changes signature everywhere on $\bar S$ and is null everywhere on $\bar S$.

As $\bar S$ is stable along $-\vec\ell$, then if ${\cal W}({\bar S}) \neq \bar S$ from point \ref{l'>l} in Result \ref{twon} $\bar S$ is strictly stable in all external directions that are non-spacelike on $\bar S\setminus {\cal W}(\bar S)$ and timelike somewhere there. In particular, $\bar S$ is strictly stable in all external timelike directions. Choosing an appropriate reference foliation, for instance with timelike hypersurfaces $\{\Sigma'_s\}$, with $\bar S\subset \Sigma'_0$, theorem \ref{th1} states that there is a MOTT adapted to $\{\Sigma'_s\}$ beyond $\bar S$ containing it, and this MOTT  will in fact be a MTT if $\bar S$ is not extremal. Any other MTT passing through $\bar S$ will be given by other external directions $\vec n'$ on $\bar S$ with $\lambda_{n'} =0$, and thus point \ref{interweave} in Result \ref{twon} states that either $\vec n'$ is different from $-\vec \ell$ only on ${\cal W}(\bar S)$, or
$$-\vec\ell -\vec n' =-\frac{n'^\mu n'_\mu}{2} \vec k
$$
changes causal character within $\bar S \setminus {\cal W}(\bar S)$, ergo $\vec n'$ has all 3 causal characters on $\bar S \setminus {\cal W}(\bar S)$. Thus, signature change always happens at some points of $\bar S$ for all MTTs that contain it.

These MTTs going beyond $\bar S$ cannot become dynamical horizons. This follows from the area law (Result \ref{area}) because the area of the MTSs $\{S'_s\}$ embedded in each $\Sigma'_s$ will be larger than $4\pi/\Lambda $ and thus point \ref{Alarger} of Result \ref{areas} implies that $\lambda_{-\ell} <0$ for all $S'_s$, so that point \ref{l'<l} in that Result \ref{areas} states that any other external direction $\vec m'$ with $\lambda_{\vec m'}=0$ must be timelike somewhere on $S'_s\setminus {\cal W}(S'_s)$. In particular the direction tangent to any of the MTTs and orthogonal to the $S'_s$ has to be timelike somewhere on $S'_s$. This behaviour must continue along the MTTs as long as the foliating surfaces remain marginally trapped and the genericity condition holds.
\square

\begin{coro}
The distinguished surface $\bar S$ does not contain any electromagnetic charges ---if the non-electromagnetic energy-momentum tensor satisfies DEC on $\bar S$.
\end{coro}
\proof From Remark \ref{rem:Tkl} $T_{\mu\nu}k^\mu \ell^\nu =0$, and if the non-electromagnetic part of $T_{\mu\nu}$ still satisfies DEC, this implies that $T^{(EM)}_{\mu\nu}k^\mu \ell^\nu =0$ for the electromagnetic energy-momentum tensor $T^{(EM)}_{\mu\nu}$. But a direct calculation (see e.g. \cite{DJR}) gives
$$
T^{(EM)}_{\mu\nu}k^\mu \ell^\nu = \frac{1}{2} \left[\left(F_{\mu\nu}k^\mu \ell^\nu \right)^2+\left((\star F)_{\mu\nu}k^\mu \ell^\nu \right)^2 \right]
$$
so that necessarily $F_{\mu\nu}k^\mu \ell^\nu|_{\bar S} = (\star F)_{\mu\nu}k^\mu \ell^\nu|_{\bar S} =0$. The electromagnetic charges enclosed by $\bar S$ are given by the integral of these quantities on $\bar S$, and thus they vanish.\square

\begin{remark}
As the distinguished surface $\bar S$ is a round sphere of constant curvature ${\cal K} =\Lambda$, it possesses several axial Killing vectors. One can then compute the angular momentum on the MTS ${\bar S}$ relative to any of these axial vectors, say $\vec \eta$, which is given by \cite{AK}
$$
J(\eta) := \int_{\bar S} \eta^A s_A = \int_{\bar S} \eta^A D_A \ln \phi_n =  \int_{\bar S} D_A (\eta^A \ln \phi_n)=0
$$
where I have used the first in \eqref{zeros} in the second equality and the fact that $\vec\eta $ is Killing on the third. Hence, all such angular momenta vanish on $\bar S$.
\end{remark}

\begin{remark}
If any one of the $S'_s$, say $S'_{s_1}$, has ${\cal W}(S'_{s_1})=\emptyset$, then (at least) some GHSs will be timelike everywhere there. To prove it, choose a {\em timelike} deformation vector $\vec n$ with $W n^\mu n_\mu =2 \lambda_{-\ell} <0$ on $S'_{s_1}$ so that, as $\lambda_{-\ell}$ is constant 
$${\rm sup}_{S'_{s_1}} \left(-W n_\mu n^\mu/2\right)={\rm inf}_{S'_{s_1}} \left(-Wn_\mu n^\mu/2\right)=-\lambda_{-\ell}$$
and thus point \ref{SupInf} in Result \ref{twon} applied to $\vec n' =-\vec\ell$ provides
$$
0\leq \lambda_n \leq 0 \Longrightarrow \lambda_n =0.
$$
Thus, some MTTs passing through $S'_{s_1}$ will be tangent to $\vec n$ and/or other nearby timelike vector fields everywhere on $S'_{s_1}$.
\end{remark}
To fix ideas, let me use ${\cal H}$ for the specific GHS that is null everywhere at $\bar S$. One may wonder what the `final fate' of ${\cal H}$ may be when proceeding towards the past with increasing areas of the foliating MTSs after having left $\bar S$ `behind'. One possibility is that eventually the MTSs become simple MOTSs. A more interesting possibility comes about if eventually a MTS $S_f$ with ${\cal W}(S_f)=S_f$ is reached. Then, one deduces that we must be in the situation of point \ref{allunsta} in Result \ref{excep}. This follows from the fact that, as explained in the proof of Result \ref{always}, the stability directions of the previous MTSs require external directions $\vec n$ closer and closer to $-\vec k$ as $W$ approaches becoming identically zero along them. After $S_f$, ${\cal H}$ may turn back to be a local timelike membrane or keep being null. In either case the `evolution' with non-decreasing area goes indefinitely towards the past. Examples of such situations are provided by the Figures 2 and 3 in \cite{Snew}. Observe, moreover, that the possibility of having other null portions of ${\cal H}$ may have happened also before reaching $\bar S$, for instance, prior to $S_0$ there may be a portion of a NEH joining the DH containing $S_0$. In this case, and for an analogous reasoning as before, the MTSs foliating these NEH portions must belong to the category of point \ref{allssta} in Result \ref{excep}. An explicit example is the spacetime of Figure 3 in \cite{Snew}.

Standard results on geodesic incompleteness \cite{HE,P,S1,SMilestone} imply the existence of a future singularity in ultra-massive spacetimes due to the existence of the future trapped surfaces. It is enough to assume, for instance, the existence of a Cauchy hypersurface $\Sigma$ containing one of the many future-trapped surfaces such that the part of $\Sigma$ external to the trapped surface is non-compact, see also section 7 in \cite{AMMS}.

\section{Discussion}\label{sec:discussion}
The results shown above may look a little counter-intuitive at first, because the `evolution' of the GHS ${\cal H}$ indefinitely towards the past implies the existence of MTSs that are unstable in all spacelike directions at the far past (or as initial data in $\scri^-$). However, one must keep in mind that we are dealing with spacetimes with positive $\Lambda$, and thus they tend to behave like de Sitter in the far past. 

The implications of the existence and properties of these ultra-massive spacetimes are reminiscent of the pioneering studies performed years ago in \cite{NYM,SNKM,BHKT,NSH} both theoretically and numerically, and also using specific explicit spacetimes---the Kastor-Traschen charged multi black hole solution \cite{KT}. The words `overmassive' \cite{BHKT} and super-critical \cite{NSH} actually already appeared there. However, the maximum bound \eqref{lim} for spatially stable MTSs is never achieved in the spacetimes analyzed in those references due to the existence of electric charge (that effectively implies a smaller bound \cite{HSN,Simon}). Nevertheless, the conclusions concerning the possibility of black hole mergers are similar: it seems that black holes with areas close to the bound \eqref{lim} will not produce a new black hole if they collide but rather something else like, for instance, an ultra-massive spacetime. The same may happen if a greedy black hole accretes too much mass from its surroundings. 

When the cosmological constant is positive, and if the spacetime admits a conformal completion with past null infinity $\scri^-$, we know \cite{Frie0} that the entire future evolution of the spacetime is encoded in initial data at $\scri^-$. Therefore, the universal properties of ultra-massive spacetimes can be characterized by some specific form of the initial data at $\scri^-$. It would be interesting to unveil the properties of such data, which might shed some light onto the definition of mass-energy at infinity when $\Lambda >0$. In any case, this way of looking at ultra-massive spacetimes resolves any puzzles concerning the anti-intuitive behaviour of the GHS ${\cal H}$ in its timelike part `after' the bound \eqref{lim} has been reached at the distinguished MTS $\bar S$, because the true future evolution of ${\cal H}$ is already determined at $\scri^-$ as long as ${\cal H}$ remains in the domain of dependence of $\scri^-$.

An important remark is that the numbers provided in the Introduction and in \cite{Snew} for the total mass of ultra-massive spherically symmetric spacetimes do not take into account the expansion of the Universe, which may change the picture in several respects, see also \cite{ZCGHSW}.

It would be very interesting to identify explicit examples of ultra-massive spacetimes without spherical symmetry ---or alternatively to disprove their existence---, for its own sake and also to gain a better intuition about their general properties and behaviours. In particular, another important question that arises is whether or not ultra-massive spacetimes may have an event horizon of an asymptotically de Sitter future infinity, or they will always lead to a universal singularity as in the examples presented in \cite{Snew}.

In this paper I have concentrated in the {\em future} case, that is to say, concerning marginally future-trapped surfaces. However, the past case could also lead to interesting conclusions. Observe that in that case the area of the MTSs along past GHS will actually increase towards the future.

\section*{Acknowledgments}
I am grateful to Marc Mars for many helpful comments and to Ken-ichi Nakao from bringing important references to my attention.
Research supported by the Basque Government grant number IT1628-22, and by Grant PID2021-123226NB-I00 funded by the Spanish MCIN/AEI/10.13039/501100011033 together with ``ERDF A way of making Europe'' . This research was carried out during a visiting professorship at Yukawa Institute for Theoretical Physics.

\end{document}